\documentclass[useAMS,usenatbib]{mn2e}
\usepackage{graphicx,amssymb,amsbsy,bm,amsmath,rotating,epsfig}
\usepackage{psfrag}
\newcommand{\be}{\begin{equation}}
\newcommand{\ee}{\end{equation}}
\newcommand{\bea}{\begin{eqnarray}}
\newcommand{\eea}{\end{eqnarray}}

\title[Model independent analysis of dark matter 
points to a particle mass at the keV scale]{Model independent 
analysis of dark matter points to a particle mass at the keV scale}
\author[H. J. de Vega, N. G. Sanchez]
{H. J. de Vega$^{1,2}$\thanks{devega@lpthe.jussieu.fr},
{N. G. Sanchez$^{2}$\thanks{Norma.Sanchez@obspm.fr}}\\
$^{1}$LPTHE, Universit\'e
Pierre et Marie Curie (Paris VI) et Denis Diderot (Paris VII),\\
Laboratoire Associ\'e au CNRS UMR 7589, Tour 24, 5\`eme. \'etage, 
Boite 126, 4, Place Jussieu, 75252 Paris, Cedex 05, France\\
$^{2}$Observatoire de Paris, LERMA. Laboratoire
Associ\'e au CNRS UMR 8112.
 \\61, Avenue de l'Observatoire, 75014 Paris, France.}
\begin{document}
\date{\today}
\pagerange{\pageref{firstpage}--\pageref{lastpage}} \pubyear{0000}

\maketitle

\label{firstpage}

\begin{abstract}
We present a {\bf model independent} analysis of dark matter (DM) both 
decoupling ultrarelativistic (UR) and non-relativistic (NR) based on the
DM phase-space density $ \mathcal{D} = \rho_{DM}/\sigma^3_{DM} $.
We derive explicit formulas for the DM particle
mass $ m $ and for the number of ultrarelativistic degrees of freedom $ g_d $
at decoupling. We find that for DM particles decoupling UR both at local 
thermal equilibrium (LTE) and out of LTE, $ m $ turns to be at the 
{\bf keV scale}. For example, for DM Majorana fermions
decoupling at LTE the mass results $ m \simeq 0.85 $ keV.
For DM particles decoupling NR,
$ \sqrt{m \; T_d} $ results in the {\bf keV scale} 
($ T_d $ is the decoupling
temperature) and the $ m $ value is consistent with the keV scale. 
In all cases, DM turns to be  {\bf cold} DM (CDM).
Also, lower and upper bounds on the DM annihilation cross-section for
NR decoupling are derived.
We evaluate the free-streaming (Jeans') wavelength and Jeans' mass: they 
result independent of the type of DM except for the DM self-gravity 
dynamics. The free-streaming wavelength today results in the  {\bf kpc range}. 
These results are based on our 
theoretical analysis, astronomical observations of dwarf spheroidal 
satellite galaxies in the Milky Way and $N$-body numerical simulations.
We analyze and discuss the results on $ \mathcal{D} $ from analytic approximate 
formulas both for linear fluctuations and the (non-linear) spherical model and from $N$-body 
simulations results. We obtain in this way upper bounds for the DM particle mass 
 which all result below the 100 keV range.
\end{abstract}
\begin{keywords}
dark matter -- keV mass scale.
\end{keywords}

\section{The dark matter particle mass}

Although dark matter was noticed seventy-five years ago \citep{zw,oo}
its nature is not yet known.

Dark matter (DM) must be non-relativistic by the time of 
structure formation ($ z < 30 $) in order to reproduce the observed small 
structure at $ \sim 2-3 $ kpc.

DM particles can decouple being 
ultrarelativistic (UR) at $ \; T_d \gg m $ or being non-relativistic (NR) 
at $ \; T_d \ll m $ where $ m $ is the mass of the dark matter particles 
and $ T_d $ the decoupling temperature. We consider in this paper 
particles that decouple {\bf at} or {\bf out} of local thermal 
equilibrium (LTE).

The DM distribution function $ F_d $ freezes out at decoupling. Therefore,
for all times after decoupling $ F_d $ coincides 
with its expression at decoupling. $ F_d $ is a function of $ T_d, \; m $ 
and the comoving momentum of the dark matter particles $ p_c $.

Knowing the distribution function $ F_d(p_c) $, we can compute physical 
magnitudes as the DM velocity fluctuations and the DM energy density. 
For the relevant times $ t $ during structure formation, when the DM 
particles are non-relativistic, we have
\be\label{flucvel}
\langle \vec{V}^2\rangle(t) = \langle \frac{\displaystyle 
\vec{p}^{\,2}_{ph}}{m^2} 
\rangle(t) = \frac{\displaystyle \int \frac{d^3 p_{ph}}{(2\pi)^3}  \;  
\frac{\vec{p}^{\,2}_{ph}}{m^2} \; 
F_d[a(t) \, p_{ph}]}{\displaystyle \int \frac{d^3p_{ph}}{(2\pi)^3} \; 
F_d[a(t) \; p_{ph}]}
\ee
where we use the physical momentum of the dark matter particles 
$ p_{ph}(t) \equiv p_c/a(t) $ 
as integration variable. The scale factor $ a(t) $ is normalized as usual,
\be \label{aoft}
a(t)= \frac1{1+z(t)} \quad , \quad a({\rm today}) = 1 \; ,
\ee 
namely, the physical momentum $ p_{ph}(t) $ coincides today with the comoving 
momentum $ p_c $.

We can relate the covariant decoupling temperature $ T_d $, the effective
number of UR degrees of freedom at decoupling $ g_d $ and the photon 
temperature today $ T_{\gamma} $ by using entropy conservation 
\citep{kt,gb,pdg}: 
\be\label{temp}
T_d = \left(\frac2{g_d}\right)^\frac13 \; T_{\gamma} \; 
, \quad {\rm where} \quad T_{\gamma} = 0.2348 \; {\rm meV} 
\ee
and $ 1 $ meV $ =  10^{-3} \; $ eV.

The DM energy density can be written as
\be \label{roDM}
\rho_{DM}(t) = g\int \frac{d^3p_{ph}}{(2\pi)^3} \; \sqrt{m^2+p^2_{ph}} ~ 
F_d[a(t) \, p_{ph}] \; ,
\ee
where $ g $ is the number of internal degrees of freedom of the DM 
particle, typically $ 1 \leq g \leq 4 $.

By the time when the DM particles are non-relativistic, the
energy density eq.(\ref{roDM}) becomes
\be \label{roDM2}
\rho_{DM}(t) = \frac{m~g}{2 \; \pi^2} \; \frac{T^3_d}{a^3(t)} \; I_2 
\equiv m \; n(t) \; , 
\ee
where
$$
I_2 \equiv \int_0^{\infty} y^2 \; F_d(y) \; dy \; ,
$$
$ n(t) $ is the number of DM particles per unit volume and 
we used as integration variable 
\be\label{varin}
y \equiv \frac{p_{ph}(t)}{T_d(t)} = \frac{p_c}{T_d} \; . 
\ee
From eq.(\ref{roDM2}) at $ t = 0 $ and from the value observed today for 
$ \rho_{DM} $ \citep{WMAP5,pdg},
\bea\label{rhoDM}
\rho_{DM} &=& \Omega_{DM} \; \rho_c = 0.228 \; \rho_c \quad , \cr \cr 
\rho_c &=& 3 \; M_{Pl}^2 \; H_0^2 = (2.518 \; {\rm meV})^4 \; ,
\eea
and $ M_{Pl}^2 = 1/[8 \, \pi \; G] $, we find the value of the DM mass:
\be\label{m}
m = \pi^2 \; \Omega_{DM} \; \frac{\rho_c}{T_{\gamma}^3} \;
\frac{g_d}{g \; I_2}=
6.986 \; \mathrm{eV} \; \frac{g_d}{g \; I_2} \; , 
\ee
where $ \rho_c $ is the critical density.

Using as integration variable $ y $ [eq.(\ref{varin})], 
eq.(\ref{flucvel}) for the velocity fluctuations, yields
\be\label{v2}
\langle \vec{V}^2\rangle(t) = \left[\frac{T_d}{m \; a(t)}\right]^2  \; 
\frac{I_4}{I_2} \; , 
\ee
where
$$
I_4 \equiv \int_0^\infty y^4  \; F_d(y)  \; dy \; .
$$
Expressing $ T_d $ in terms of the CMB temperature today according to
eq.(\ref{temp}) gives for the one-dimensional velocity dispersion,
\bea\label{sigdm}
&\sigma_{DM}(z) = \sqrt{\frac13 \; \langle \vec{V}^2 \rangle(z)} = 
\displaystyle \frac{2^{\frac13}}{\sqrt3} \; \frac{1+z}{g^{\frac13}_d} \; 
\frac{T_{\gamma}}{m} \; \sqrt{\frac{I_4}{I_2}}=& \\ \cr
& =0.05124 \; \displaystyle 
\frac{1+z}{g^{\frac13}_d} \; \frac{\mathrm{keV}}{m}  \;
\left[\frac{I_4}{I_2}\right]^{\frac12} \; \frac{\mathrm{km}}{\mathrm{s}}& \; \; .
\eea
It is very useful to consider the phase-space density invariant under the 
universe expansion \citep{nos1,hogan,madsen,madsen2}
\be \label{defD}
\mathcal{D}(t)  \equiv \frac{n(t)}{\langle \vec{P}^2_{ph}(t) 
\rangle^\frac32} \buildrel{\rm non-rel}\over= 
\frac1{3 \; \sqrt3  \; m^4} \; \frac{\rho_{DM}(t)}{\sigma^3_{DM}(t)}
\quad , 
\ee
where we consider the relevant times $ t $ during structure formation when
the DM particles are non-relativistic.
$ \mathcal{D}(t) $ is a \emph{constant} in absence of self-gravity. In the
non-relativistic regime $ \mathcal D(t) $ can only {\bf decrease} by
collisionless phase mixing or self-gravity dynamics \citep{theo,theo2}.

We derive a useful expression for the phase-space density $ {\mathcal D} $
from eqs.(\ref{roDM2}), (\ref{sigdm}) and (\ref{defD}) with the result 
\be \label{Dex} 
\displaystyle \mathcal{D} = \frac{g}{2 \; \pi^2}
\frac{I_2^{\frac52}}{I_4^{\frac32}}\; , 
\ee

\noindent
Observing dwarf spheroidal satellite galaxies in the Milky Way (dSphs) 
yields for the phase-space density today \citep{gilmore,gilmore2}:
\be\label{gil}
\frac{\rho_s}{\sigma^3_s} \sim 5\times 10^3 ~
\frac{\mathrm{keV}/\mathrm{cm}^3}{\left( \mathrm{km}/\mathrm{s}
\right)^3} = (0.18 \;  \mathrm{keV})^4 \; .
\ee
The precision of these results is about a factor $10$.

After the radiation dominated era the phase-space density reduces 
by a factor that we call $ Z $ 
\be\label{F1}
\mathcal{D}(0) = \frac1{Z} \; \mathcal{D}(z \sim 3200)
\ee
Recall that $ \mathcal{D}(z) = \rho_{DM}/(3 \; \sqrt3  \; m^4 \; \sigma^3_{DM}) $ 
[according to eq.(\ref{defD})] is independent of $ z $ for $ z \gtrsim 3200 $
since density fluctuations were $ \lesssim 10 ^{-3} $ before the matter dominated era \citep{dod}.

The range of values of $ Z $ (which is necessarily  $ Z > 1 $) is analyzed in detail
in sec. \ref{snZel} below.

We can express the phase-space density today from eqs.(\ref{defD}) and (\ref{gil}) as
\be\label{D0}
\mathcal{D}(0) = \frac1{3 \; \sqrt3  \; m^4} \; \frac{\rho_s}{\sigma^3_s} \; .
\ee
Therefore, eqs.(\ref{defD}), (\ref{F1}) and (\ref{D0}) yield,
\be\label{F}
\frac{\rho_s}{\sigma^3_s} = \frac1{Z} \; \frac{\rho_{DM}}{\sigma^3_{DM}}(z \sim 3200) \; ,
\ee
where $ \rho_{DM}/\sigma^3_{DM}(z \sim 3200) $ follows from eqs.(\ref{defD}) and (\ref{Dex}),
\be\label{denhoy}
\frac{\rho_{DM}}{\sigma^3_{DM}}(z \sim 3200) = \frac{3 \; \sqrt3  \; m^4}{2 \; \pi^2} \;
g \;\frac{I_2^{\frac52}}{I_4^{\frac32}} \; . 
\ee
We can express $ m $ from eqs.(\ref{gil})-(\ref{denhoy}) in terms of 
$ \mathcal{D} $ and observable quantities as
\bea\label{mD}
&&m^4 = \frac{Z}{3 \; \sqrt3} \; \frac{\rho_s}{\mathcal{D} \; \sigma^3_s}
= \frac{2 \; \pi^2}{3 \; \sqrt3} \; \frac{Z}{g} \; 
\frac{\rho_s}{\sigma^3_s} \; \frac{I_4^{\frac32}}{I_2^{\frac52}} \; , \\ \cr
&& m = 0.2504 \; \left(\frac{Z}{g}\right)^\frac14 \; \; 
\frac{I_4^{\frac38}}{I_2^{\frac58}} \; 
\mathrm{keV} \;  . \label{mDn}
\eea
Combining this with eq.(\ref{m}) for $ m $ we obtain the number of 
ultrarelativistic degrees of freedom at decoupling as
\bea\label{gdD}
&&g_d = \frac{2^\frac14}{3^\frac38 \; \pi^\frac32} \; 
\frac{g^\frac34}{\Omega_{DM}} \; \frac{T_{\gamma}^3}{\rho_c} \; 
\left(\frac{Z \; \rho_s}{\sigma^3_s}\right)^\frac14 \;
\left[I_2 \; I_4\right]^{\frac38} \cr \cr
&&= 35.96 \; Z^\frac14 \; g^\frac34 \; \left[I_2 \; I_4\right]^{\frac38} \; .
\eea

\noindent
If we assume that dark matter today is a self-gravitating gas in thermal 
equilibrium described by an isothermal sphere solution of the Lane-Emden 
equation, the relevant quantity characterizing the dynamics is the 
dimensionless variable \citep{gas,gas2} 
\be\label{L} 
\eta = \frac{G \; m^2 \; N}{L \; T} = 
\frac23 \; G \; L^2 \; \frac{\rho_s}{\sigma_s^2} \; ,
\ee
which is bound to be $ \eta \lesssim 1.6 $ to prevent the
gravitational collapse of the gas \citep{gas,gas2}. Here $ V = L^3 $ stands 
for the volume occupied by the gas, $ N $ for the number of particles, 
$ G $ for Newton's constant and $ T = \frac32 \; m \; \sigma^2 $ is the 
gas temperature. (The length $ L $ is similar to the so-called King 
radius \citep{bt}. Notice however that the King radius follows from the singular 
isothermal sphere solution while $ L $ is the characteristic size of a 
stable isothermal sphere solution \citep{gas,gas2}.)

\medskip

The compilation of recent photometric and kinematic data from ten
Milky Way dSphs satellites \citep{gilmore,gilmore2} yields values for the one
dimensional velocity dispersion $ \sigma_s $ and the radius
$ L $  in the ranges
\be\label{gilsL}
0.5~\mathrm{kpc} \leq L \leq 1.8 ~\mathrm{kpc} \quad , \quad
6.6~\mathrm{km}/\mathrm{s} \leq \sigma_s \leq 11.1 ~
\mathrm{km}/\mathrm{s} \; . 
\ee
Combining eq.(\ref{defD}), eq.(\ref{F}) and (\ref{L}) yields the
explicit expression for the DM particle mass,
\bea\label{meta}
&m^4 \sim \displaystyle \frac1{2 \; \sqrt3 \; G} \; \eta \; 
\frac{Z}{L^2 \; {\mathcal D} \; \sigma_s}
=  \displaystyle \frac{4 \; \pi}{\sqrt3} \; M_{Pl}^2 \; \eta \; \frac{Z}{L^2 \; 
{\mathcal D} \; \sigma_s}& \cr \cr
& = 0.5279 \; 10^{-4} \; \displaystyle
\frac{\eta \;Z}{\mathcal D} \; 
\frac{10 \; \mathrm{km}/\mathrm{s}}{\sigma_s} \; 
\left(\frac{\rm kpc}{L}\right)^2 \; ({\rm keV})^4 &\; .
\eea
This formula provides an expression for the DM particle mass 
{\bf independent} of eq.(\ref{mD}). We shall see below that 
eqs.(\ref{mD}) and (\ref{meta}) yield similar results. 

We investigate in the subsequent sections the cases where DM particles 
decoupled UR or NR both at LTE and out of LTE. We compute there $ m $ and
$ g_d $ explicitly in the different cases according to the general 
formulas eqs.(\ref{mDn}), (\ref{gdD}) and (\ref{meta}).

\subsection{Jeans'  (free-streaming)  wavelength and Jeans' mass}\label{jeans}

It is very important to evaluate the Jeans' length and Jeans' mass in the 
present context \citep{gb,gilbert,bs}. The Jeans' length is analogous to 
the free-streaming wavelength. The free-streaming wavevector is the largest 
wavevector exhibiting gravitational instability and characterizes the 
scale of suppression of the DM transfer function $ T(k) $ \citep{egil}.

The physical free-streaming wavelength can be expressed as \citep{gb,egil}
\be\label{fs}
\lambda_{fs}(t) = \lambda_J(t) = \frac{2 \; \pi}{k_{fs}(t)}
\ee
where $ k_{fs}(t) = k_J(t) $ is the  physical free-streaming wavenumber given by
\be\label{kfs}
k_{fs}^2(t) = \frac{4 \; \pi \; G \; \rho_{DM}(t)}{\langle 
\vec{V}^2\rangle(t)}= \frac32 \; [1+z(t)] \; \frac{H_0^2 \; 
\Omega_{DM}}{\langle \vec{V}^2\rangle(0)} \;  \; .
\ee
where we used that $ \rho_{DM}(t) = \rho_{DM}(0) \; (1+z)^3 $ and 
eq.(\ref{rhoDM}). 

We obtain the primordial DM dispersion velocity $ \sigma_{DM} $ from eqs. (\ref{roDM2}), 
(\ref{rhoDM}) and (\ref{F}),
\be\label{sig2}
\sqrt{\frac13 \; \langle \vec{V}^2\rangle(0)} = 
\sigma_{DM} = \left(3 \, \; M_{Pl}^2 \; H_0^2 \; \Omega_{DM} \; \frac1{Z} 
\; \frac{\sigma^3_s}{\rho_s} 
\right)^{\frac13}
\ee
This expression is valid for {\bf any kind} of DM particles.
Inserting eq.(\ref{sig2}) into eq.(\ref{kfs}) yields for the physical free-streaming
wavelength
\bea\label{fs2}
\lambda_{fs}(z) &=& \frac{2 \, \sqrt2 \, \pi}{\Omega_{DM}^\frac16} \;
\left(\frac{3 \; M_{Pl}^2}{H_0}\right)^\frac13 \;
\left(\frac{\sigma_s^3}{Z \; \rho_s}\right)^\frac13 \; \frac1{\sqrt{1+z}} =
\cr \cr
&=&\frac{16.3}{Z^\frac13} \;  \; \frac1{\sqrt{1+z}} \; \; {\rm kpc} \; . 
\eea
where we used $ 1 $ keV $ = 1.563738 \; 10^{29} \; ({\rm kpc})^{-1} $.

Notice that $ \lambda_{fs} $ and therefore $ \lambda_J $ turn to be {\bf independent} 
of the nature of the DM particle except for the factor $ Z $.

The approximated analytic evaluations in sec. \ref{secdos} together with the results
of $N$-body simulations \citep{numQ1,numQ2,numQ3,numQ4,numQ5,numQ6}
indicate that for dSphs $ Z $ is in the range 
$$ 
1 < Z < 10000 \; .
$$ 
Therefore, $ 1 < Z^\frac13 < 21.5 $ and the free-streaming wavelength
results in the range
$$
0.757 \; \frac1{\sqrt{1+z}} \; {\rm kpc} <\lambda_{fs}(z) <
16.3 \; \frac1{\sqrt{1+z}} \; {\rm kpc} \; .
$$
These values at $ z = 0 $ are consistent with the $N$-body simulations 
reported in \citet{gao} and are of the order of the small DM structures 
observed today \citep{gilmore,gilmore2}.

The Jeans' mass is given by
\be
M_J(t) = \frac43 \; \pi \; \lambda_J^3(t) \; \rho_{DM} (t) \; .
\ee
and provides the smallest unstable mass by gravitational collapse 
\citet{kt,gb}. Inserting here eq.(\ref{roDM2}) for the DM density and 
eq.(\ref{fs2}) for $ \lambda_J(t) = \lambda_{fs}(t) $ yields 
\bea \label{masJ}
M_J(z) &=& 192 \, \sqrt2 \, \pi^4 \; \sqrt{\Omega_{DM}} \; M_{Pl}^4 \; H_0 \; 
\frac{\sigma_s^3}{Z \; \rho_s} \; (1+z)^\frac32  =\cr \cr
&=&\frac{0.4464}{Z} \; 10^7 \; M_{\odot} \; (1+z)^\frac32 \; .
\eea
Taking into account the $Z$-values range yields
$$
0.4464 \; 10^3 \; M_{\odot} < M_J(z)\; (1+z)^{-\frac32} < 0.4464 \; 10^7  \; 
\; M_{\odot} \; .
$$
This gives masses of the order of galactic masses
$ \sim 10^{11} \; M_{\odot} $ by the beginning of the MD era $ z \sim 3200 $.
In addition, the comoving free-streaming wavelength scale by $ z \sim 3200 $ 
$$
3200 \times \lambda_{fs}(z \sim 3200) \sim 100 \; {\rm kpc}  \; ,
$$
turns to be of the order of the galaxy sizes today.

\section{The phase-space density $ \mathcal{D} $ from analytic approximation
methods and from $N$-body simulations}\label{secdos}

We analytically derive here formulas for the reduction factor $ Z $ defined by eq.(\ref{F1}) 
in the  linear approximation and in the spherical model.
The results obtained (see Table I) are in fact {\bf upper bounds} for $ Z $.
We then analyze the results on  $ \mathcal{D} $ from $N$-body simulations.

\subsection{Linear perturbations}\label{seclin}

The simplest calculation of $ \mathcal{D} $ follows by considering linear perturbations
around the homogeneous distribution $ \rho_{DM}(z) $ as
\be
\rho = \rho_{DM}(z)\left[ 1 + \delta(z,k) \right] \; ,
\ee
We have $ \rho_{DM}(z) = \rho_{DM}(0) \; (1+z)^3 $ and in a matter dominated universe
\be\label{delf}
\delta(z,k) \sim \delta_i \; \frac{1+z}{1+z_i} \; .
\ee
The peculiar velocity in the MD universe behaves as \citep{dod},
\be
v \sim a \; H \; \delta \sim 1/\sqrt{1+z} \; .
\ee
We can thus relate the phase-space density at redshift $ z $ 
$ \mathcal{D}(z)\sim \rho/v^3 $ eq.(\ref{defD})
with the phase-space density at redshift $ z_i $ as,
\be\label{Dfl}
\mathcal{D}(z)\sim \mathcal{D}(z_i) \; \left(\frac{1+z}{1+z_i}\right)^{\frac92} \; .
\ee
Since the linear approximation is valid for $ |\delta|^2 \ll 1 $,
we find from eq.(\ref{delf}) that eq.(\ref{Dfl}) applies in the redshift 
range $ (z_i, \, z) $ where
\be\label{valid}
1+z \gg (1+z_i) \; \delta_i \; .
\ee
We can apply eq.(\ref{Dfl}) to relate $ \mathcal{D}(z) $ at equilibration
($ z=z_i \simeq 3200 $) and $ \mathcal{D}(z) $ at the beginning of structure
formation $ z \sim 30 $, since  $ \delta_i \sim 10^{-3} $ as a scale average of the density fluctuations at
the end of the RD dominated era \citep{dod} and eq.(\ref{valid}) is satisfied.
We thus obtain from  eq.(\ref{Dfl}) in the linear approximation:
\be\label{vZlin}
\frac{\mathcal{D}(z\simeq 3200)}{\mathcal{D}(z\simeq 30)} \sim  1.3 \times 10^9 \; .
\ee
Notice that eq.(\ref{valid}) does not hold for $ z \sim 0 $.
Therefore, in order to evaluate $ Z $, we should combine 
the linear approximation result eq.(\ref{vZlin}) for $ 30 \lesssim z \lesssim 3200 $
with the results of $N$-body simulations for $ 0 \lesssim z \lesssim 30 $.
This is done in sec. \ref{discuZ} to obtain upper bounds for $ Z $.

\begin{table*}
 \centering
 \begin{minipage}{140mm}
  \begin{tabular}{|c|c|c|} \hline  
Approximation used & Upper limit on $ Z $   &  Upper limit on $ m \simeq 0.5 \; Z^\frac14 \; $ keV \\
\hline 
  Linear fluctuations &  $ \sim 1.3 \times 10^{11} $ & $ 96 $ keV \\
\hline 
Spherical Model &  $ \sim 1.29 \times \delta_i^{-\frac32} \simeq 4.1 \times 10^4 $ & $ 7.1 $ keV\\
\hline   
\end{tabular}
\caption{Upper bounds for the $Z$-factor [defined by eq.(\ref{F1})] and for the mass
of the DM particle obtained for two different approximation methods.
Notice that only the spherical model takes into account non-linear self-gravity effects.
The mass $ m $ {\bf mildly} depends on $ Z $ through the power $ 1/4 $. In any case $ m $
results in the keV range.}
\end{minipage}
\end{table*}

\subsection{The spherical model}\label{secesf}

Let us now consider the spherical model where particles only move in the radial
direction but where the non-linear evolution is exactly solved \citep{FG,EB,PJEP,TP}.
The proper radius of the spherical shell obeys the equation
\be\label{esf}
{\ddot R} = -\frac{G \; M}{R^2}
\ee
where $ G $ is the gravitational constant and $ M $ the (constant) mass enclosed by the shell.
Eq.(\ref{esf}) can be solved in close form with the solution \citep{EB,PJEP} 
\bea\label{solesf}
&& t = \frac{3 \, t_i}{4 \, \delta_i^\frac32} \; (\theta-\sin \theta) \; , \cr \cr
&& R = \frac{R_i}{2 \, \delta_i} \; (1 - \cos \theta) \quad , \quad
\frac{2 \, R_i^3}{9 \; t_i^2} = G \; M \; , \cr \cr
&& {\dot R} =  \frac{2 \, R_i \; \sqrt{\delta_i}}{3 \, t_i} \; 
\frac{\sin \theta}{1 - \cos \theta}\; , \cr \cr
&& 1+z = (1+z_i) \; \delta_i \; \left(\frac43 \right)^\frac23 \; 
\frac1{(\theta-\sin \theta)^\frac23} \; , \cr \cr
&& \rho = \rho_{DM}(z) \; \frac92 \; 
\frac{(\theta-\sin \theta)^2}{(1 - \cos \theta)^3} \; , 
\eea
Here, $ R_i $ and $ z_i $ are the radius and the redshift at the initial time $ t_i $ 
and $ \theta $ is an auxiliary time dependent parameter.

Choosing the initial time by equilibration with  $ z_i \gg 1 $ we have
$ \theta_i \ll 1 $ and we find from eqs.(\ref{solesf}),
\be\label{inic}
\theta_i = 2 \; \sqrt{\delta_i} \quad , \quad
{\dot R}(t_i) = \frac{2 \, R_i}{3 \, t_i} \quad , \quad \rho_i = \rho_{DM}(z_i) \; .
\ee
The spherical shell reaches its maximum radius of expansion $ R_m = R_i/\delta_i $
at $ \theta = \pi $ and then it turns around and collapses to a point at $ \theta = 2 \, \pi $.
However, well before that, the approximation that matter only moves radially
and that random particle velocities are small will break down. Actually,
the DM relaxes to a virialized configuration where the velocity and 
the virial radius follow from the virial theorem \citep{TP}
\be\label{viri}
v^2 = \frac{6 \, G \; M}{5 \, R_m} \quad , \quad R_v = \frac12 \;  R_m \; .
\ee 
We can now compute the initial phase-space density (at  $ z_i $) and
the phase-space density at virialization. We get at $ z_i $ from eqs.(\ref{defD})
and (\ref{inic})
\be
\mathcal{D}_i = \frac{\rho_{DM}(0)}{3 \; \sqrt3  \; m^4} \; (1+z_i)^3 \;
\left(\frac{3 \, t_i}{2 \, R_i}\right)^3 \; , 
\ee
and at virialization for $ \theta = 2 \, \pi $ from eqs.(\ref{defD}) and (\ref{viri}) 
\be
\mathcal{D}_v = \frac{\rho_{DM}(0)}{3 \; \sqrt3  \; m^4} \; (1+z_i)^3 \;
\left(\frac{t_i}{R_i}\right)^3 \; \frac{32}{9 \, \pi^2} \; 
\left(\frac{15}4 \; \delta_i\right)^\frac32 \; \; .
\ee
Therefore, the $ Z $ factor in the spherical model takes the value
$$
Z = \frac{\mathcal{D}_i}{\mathcal{D}_v} = \frac{9 \, \pi^2}{32} \; 
\left(\frac3{5 \, \delta_i}\right)^\frac32 = \frac{1.29009}{\delta_i^\frac32}
\; .
$$
Setting $ \delta_i \sim 10^{-3} $ as a scale average of the density fluctuations at
the end of the RD dominated era \citep{dod} yields
\be\label{vZesf}
Z \sim 4.08 \times 10^4 \; .
\ee
The spherical model approximates the evolution as a purely radial
expansion followed by a radial collapse. Since no transverse motion
is allowed neither mergers, the spherical model result for $ Z $ eq.(\ref{vZesf}) is 
actually an upper bound on $ Z $.

\subsection{The phase-space density $ \mathcal{D} $  from $N$ body simulations}\label{snZel}

The phase-space density $ \mathcal{D}(z) $ is invariant under the universe expansion
except for the self-gravity dynamics that diminishes $ \mathcal{D}(z) $ in its evolution
\citep{theo,theo2}. Numerical simulations show that  $ \mathcal{D}(z) $ decreases sharply during 
phases of violent mergers followed by quiescent phases \citep{numQ1,numQ2,numQ3,numQ4,numQ5,numQ6}.
$ \mathcal{D}(z) $ decreases at these violent phases by a factor of the order 
$ \gtrsim 1 $. (See fig. 3 in  \citet{numQ1}, fig. 1 in \citet{numQ2}, fig. 6 in
\citet{numQ3} and fig. 5 in \citet{numQ6}). These sharp decreasings of $ {\cal D} $
are in agreement with the linear approximation of sec. \ref{seclin} as we show below.

A succession of several violent phases happens during the structure formation 
stage ($ z \lesssim 30 $). Their cumulated effect together with the
evolution of $ \mathcal{D} $ for $ 3200 \gtrsim z \gtrsim 30 $ produces a range of values of the 
$ Z $ factor which we can conservatively estimate
on the basis of the $N$-body simulations results
\citep{numQ1,numQ2,numQ3,numQ4,numQ5,numQ6} and the approximation results 
eqs.(\ref{vZlin}) and (\ref{vZesf}). This gives a range of values 
$ 1 < Z < 10000 $ for dSphs.

Indeed, more accurate analysis of $ N $ body simulations
should narrow this range for $ Z $ which depends on the type and size of the galaxy considered. 

The dSphs observations, for which the best 
observational data are available, take mostly
into account the cores of the structures since the dSphs have been stripped of their 
external halos. Hence, the observed values of  $ \rho_s/\sigma^3_s $ may be higher
than the space-averaged valued represented by the right hand side of
eqs.(\ref{F1}) and (\ref{F}). Higher values for $ \rho_s/\sigma^3_s $ 
correspond to {\bf lower} values for $ Z $. 

\medskip

The approximate formula eq.(\ref{Dfl}) indicates a sharp decrease of the 
phase-space density with the redshift. This sharp decreasing is in qualitative agreement with the 
simulations in the violent phases \citet{numQ1,numQ2,numQ3,numQ4,numQ5,numQ6}. 

\subsection{Sinthetic discussion on the evaluation of $ Z $ and its upper bounds.}\label{discuZ}

The DM particle mass scale is set by the phase-space density for dSphs eq.(\ref{gil}).
Those galaxies are particularly dense and exhibit larger values for $ \rho_s/\sigma^3_s $
than spiral galaxies. Since the primordial phase-space density is an universal
quantity only depending on cosmological parameters, the $ Z $-factor must be galaxy
dependent, larger for spiral galaxies than for dSphs.

\medskip

We want to stress that the values of the relevant quantities $ m $ and $ g_d $ 
are {\bf mildly} affected by the uncertainity of $ Z $ through the factor $ Z^\frac14 $ 
[see eqs.(\ref{mDn})-(\ref{gdD})].

\medskip

Eqs.(\ref{Dfl}) provides an extreme high estimate for the decrease of 
$ \mathcal{D} $ and hence an extreme high estimate for $ Z $.
The $N$-body simulations show that the violent decrease of $ \mathcal{D} $ 
is restricted to a factor of order one at each violent phase
\citep{numQ1,numQ2,numQ3,numQ4,numQ5,numQ6}.

\medskip

In summary, the linear approximation suggests a reduction of  
$ \mathcal{D} $ at each violent phase by a factor $ \gtrsim 1 $ 
while such approximation is valid. Succesive violent phases can 
reduce $ \mathcal{D} $ by a factor up to $ \sim 10 $ in the range $ 0 \lesssim z \lesssim 30 $
as shown in the simulations \citet{numQ1,numQ2,numQ3,numQ4,numQ5,numQ6}.

\medskip

Combining the approximate decrease of $ \mathcal{D}(z) $ given by eq.(\ref{vZlin}) with
an upper bound of a decrease by a factor $ \sim 100 $ for the interval $ 0 \lesssim z \lesssim 30 $
yields in the linear approximation the upper bound
\be
 Z <  1.3 \times 10^{11} \quad  ,
\ee
and we have eq.(\ref{vZesf}) for $ Z $ in the spherical model. The fact that $ Z $ 
in the spherical model turns to be several orders of magnitude
{\bf below} the $ Z $ value in the linear approximation arises from the fact
that the spherical model does include non-linear effects
and it is therefore somehow {\bf more reliable} than the linear approximation.

\medskip

The range $ 1 < Z < 10000 $  for dSphs from $N$-body simulations corresponds to realistic initial 
conditions in the simulations.

\medskip

The evolutions in the two approximations considered (see Table I) are simple
spatially isotropic expansions, followed by a collapse in the case of the spherical model. There is no
possibility of non-radial motion neither of mergers in these approximations
contrary to the case in $N$-body simulations. 
For such reasons, the $Z$-values in Table I are {\bf upper bounds} to the true values of $ Z $ in galaxies. 
The largest bound on $ Z $ yield DM particle masses below $ \sim 100 $ keV. Moreover,
the  {\bf more reliable} spherical model yields $ 7.1 $ keV as upper bound
for the DM particle mass.

\medskip

In summary, with {\bf realistic} initial conditions $ \mathcal{D} $
will not decrease more than $ \lesssim 10000 $ and it is therefore 
{\bf fair} to assume that $ Z < 10000 $ for dSphs.

\section{DM particles decoupling being ultrarelativistic}

\subsection{Decoupling at Local Thermal Equilibrium (LTE)}\label{dlte}

If the dark matter particles of mass $ m $ decoupled at a temperature 
$ T_d \gg m $ their freezed-out distribution function only depends on 
$$ 
\frac{p_c}{T_d} = \frac{p_{ph}(t)}{T_d(t)} , \quad {\rm where} \quad
T_d(t) \equiv \frac{T_d}{a(t)} \; .
$$ 
That is, the distribution function for
dark matter particles that decoupled in thermal equilibrium takes the form
$$
F_d^{equil} \left[\frac{p_{ph}(t)}{T_d(t)}\right] = 
F_d^{equil}\left[\frac{p_c}{T_d}\right] \; ,
$$
where $ F_d^{equil} $ is a Bose-Einstein or Fermi-Dirac distribution 
function:
\be \label{fdbe}
F_d^{equil}[p_c] = \frac1{\exp[\sqrt{m^2+p^2_c}/T_d]\pm 1} \; .
\ee
Notice that for eq.(\ref{fdbe}) in this regime:
$$
\frac{\sqrt{m^2+p^2_c}}{T_d} \buildrel{T_d \gg m}\over= y + 
{\cal O}\left(\frac{m^2}{T_d^2}\right) \; .
$$
where $ y $ is defined by  eq.(\ref{varin}) and we can use as distribution functions
\be\label{disUR}
F_d^{equil}(y) = \frac1{e^y \pm 1} \; .
\ee
Using eqs.(\ref{m}) and (\ref{fdbe}), we find then for Fermions and 
for Bosons decoupling at LTE
\be\label{mequil}
m = \frac{g_d}{g} \; 
\left\{  \begin{array}{l} 
3.874 \; \mathrm{eV} \; 
~~\mathrm{Fermions} \\
2.906 \; \mathrm{eV} \; ~~\mathrm{Bosons} 
\end{array} \right. \; .
\ee
We see that for DM that decoupled at the Fermi scale: $ T_d \sim 100 $ GeV
and $ g_d \sim 100, \;  m $ results in the keV scale 
as already remarked in \citet{bs,bstpp,bst}.
DM particles may decouple earlier with  $ T_d > 100 $ GeV but
$ g_d $ is always in the hundreds even in grand unified theories 
where $ T_d $ can reach the GUT energy scale.
Therefore, eq.(\ref{mequil}) {\bf strongly suggests} that the mass
of the DM particles which decoupled UR in LTE is in the {\bf keV scale}.

It should be noticed that the Lee-Weinberg \citep{LW,SK,VDZ} lower bound
as well as the Cowsik-McClelland \citep{cow} upper bound 
follow from eq.(\ref{m}) as shown in \citet{nos1}.

\medskip

Computing the integrals in eq.(\ref{Dex}) with the distribution functions
eq.(\ref{fdbe}) yields for DM decoupling UR in LTE
\be\label{LTDD}
\mathcal{D} = g \; \left\{  \begin{array}{l}
\frac1{4 \; \pi^2} \; \sqrt{\frac{\zeta^5(3)}{15 \; \zeta^3(5)}}
= 1.9625\times 10^{-3}~~\mathrm{Fermions} \\
\frac1{8 \; \pi^2}\; \sqrt{\frac{\zeta^5(3)}{3 \; \zeta^3(5)}}
=3.6569\times 10^{-3}~~\mathrm{Bosons} 
\end{array} \right.
\ee
where $ \zeta(3) = 1.2020569\ldots $ and $ \zeta(5) = 1.0369278 \ldots $.

\medskip

Inserting the distribution function eq.(\ref{disUR}) into eqs.(\ref{mD}) 
and (\ref{gdD}) for $ m $ and $ g_d $, respectively, we obtain 
\bea\label{mgdeq}
 m &=& \left(\frac{Z}{g}\right)^\frac14 \; \mathrm{keV} \; 
\left\{\begin{array}{l}
         0.568~~~\mathrm{Fermions} \\
              0.484~~~\mathrm{Bosons}      \end{array} \right. \quad , \cr \cr
 g_d &=& g^\frac34 \; Z^\frac14 \; \left\{\begin{array}{l}
         155~~~\mathrm{Fermions} \\
              180~~~\mathrm{Bosons}      \end{array} \right. \; . 
\eea
Since $ g = 1-4 $, for DM particle decoupling at LTE, we see from 
eq.(\ref{mgdeq}) that $ g_d > 100 $ and thus, the DM particle should 
decouple for $ T_d > 100 $ GeV. Notice that $ 1 < Z^\frac14 < 10 $ for 
$ 1 < Z < 10000 $.

\medskip

A further estimate for the DM mass $ m $ follows by
inserting eq.(\ref{LTDD}) for $ \mathcal{D} $ in eq.(\ref{meta}) 
\be\label{metae}
m \sim \left[\frac{\eta \; Z}{g} \; 
\frac{10 \; \mathrm{km}/\mathrm{s}}{\sigma_s}\right]^\frac14 \;
\sqrt{\frac{\rm kpc}{L}} \; {\rm keV} \;  \left\{\begin{array}{l}
     0.405 ~~\mathrm{Fermions} \\
     0.347 ~~\mathrm{Bosons} 
 \end{array} \right.  \; .
\ee
Taking into account the observed values for $ \sigma_s $ and $ L $ from 
eq.(\ref{gilsL}) and the fact that $ \eta \lesssim 1.6 , \; g 
\simeq 1-4, \; 1 < Z^\frac14 < 10 $, eq.(\ref{metae}) gives again a mass
$ m $ in the keV scale as in eq.(\ref{mgdeq}). Both equations 
(\ref{mgdeq}) and (\ref{metae}) yield a mass larger in 17\% for the 
fermion than for the boson.

\medskip

\begin{figure}
\begin{turn}{-90}
\psfrag{"fd.dat"}{Fermions}
\psfrag{"be.dat"}{Bosons}
\includegraphics[height=9.cm,width=9.cm]{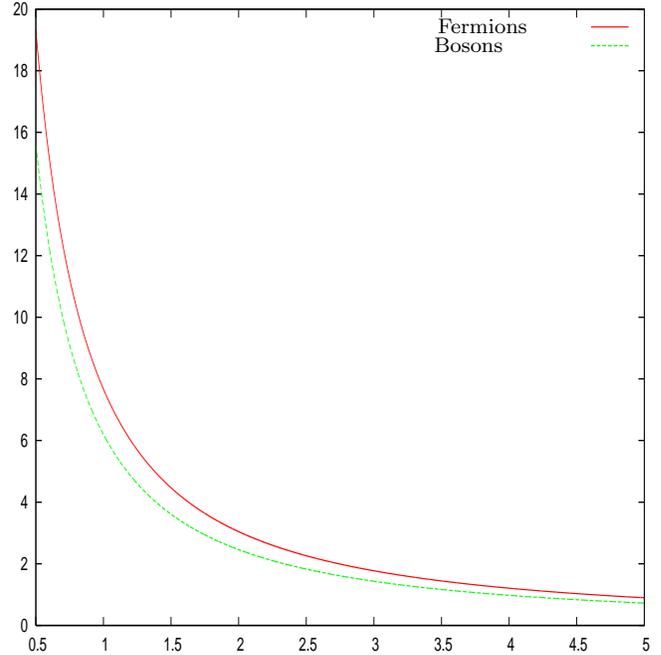}
\end{turn}
\caption{The free-streaming wavelength today $ \lambda_{fs}(0) $ in kpc vs. 
the DM particle mass $ m $ in keV times $ g^{\frac14} $ for ultrarelativistic decoupling 
at LTE according to eq.(\ref{fsfdbe}).} 
\label{ffs}
\end{figure}

We can express the free-streaming wavelength as a function of the DM particle mass
from eqs.(\ref{fs2}) and (\ref{mgdeq}) with the result,
\be\label{fsfdbe}
\lambda_{fs}(z) = \left(\frac{\rm keV}{m}\right)^\frac43 \; \frac{\rm kpc}{g^\frac13} \;
\frac1{\sqrt{1+z}} \; \left\{\begin{array}{l}
     7.67 ~~\mathrm{Fermions} \\
     6.19 ~~\mathrm{Bosons} 
 \end{array} \right.  \; .
\ee
We display in fig. \ref{ffs} $ \lambda_{fs}(0) $ in kpc vs. $ m \; g^{\frac14} $ in keV.

\subsection{Decoupling out of LTE}\label{dflte}

In general, for DM decoupling out of equilibrium, the DM particle d
istribution function takes the form
\be \label{fdf}
 F_d(p_c) = F_d\left(\frac{p_c}{T_d};\frac{m}{T_d};\ldots\right) 
\ee
Typically, thermalization is reached by the mixing of the particle modes
and scattering between particles that redistributes the particles in phase
space: the larger momentum modes are populated
by a \emph{cascade} whose front moves towards the ultraviolet
akin to a direct cascade in turbulence, leaving in its wake a state
of nearly LTE but with a {\it lower} temperature than that of
equilibrium \citep{DdV,DdV2}. Hence, in the case the dark matter particles are 
not yet at thermodynamical equilibrium at decoupling, their momentum 
distribution is expected to be peaked at smaller momenta
since the ultraviolet cascade is not yet completed \citep{DdV,DdV2}. The 
freezed-out of equilibrium distribution function can be then written as
\be \label{fuera}
F_d^{out~of~LTE} (p_c)= 
F_0 \; F_d^{equil}\left[\frac{a(t) \; p_{ph}(t)}{\xi \; T_d}\right] \; 
\theta(p_c^0 - p_c) \; ,
\ee
where $ \xi = 1 $ at thermal equilibrium and $ \xi < 1 $ {\bf before}  
thermodynamical equilibrium is attained. $ F_0 \sim 1 $ is a normalization 
factor and $ p_c^0 $ cuts the spectrum in the UV region not yet reached by 
the cascade.

\begin{figure}
\begin{turn}{-90}
\psfrag{"xf.dat"}{$ X_+(s) $ vs. $ s $}
\psfrag{"xb.dat"}{$ X_-(s) $ vs. $ s $}
\psfrag{"wf.dat"}{$ W_+(s) $ vs. $ s $}
\psfrag{"wb.dat"}{$ W_-(s) $ vs. $ s $}
\includegraphics[height=9.cm,width=9.cm]{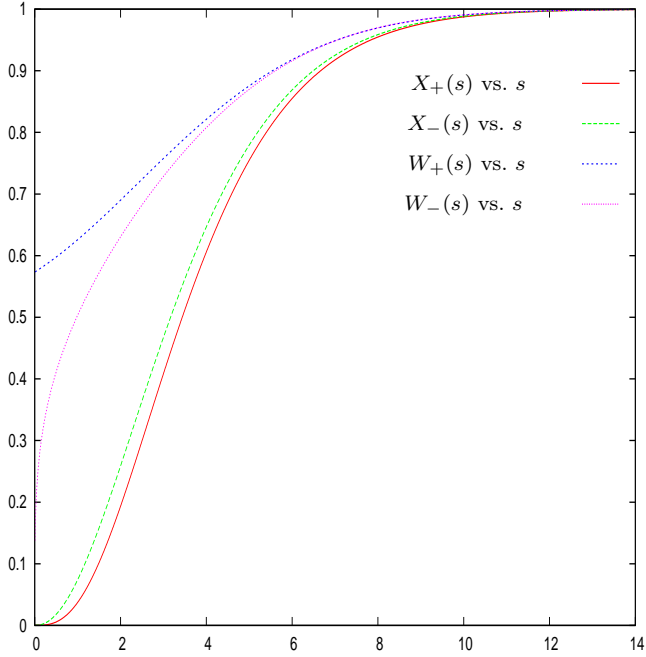}
\end{turn}
\caption{$ X_+(s) , \;  X_-(s) , \;
 W_+(s) $ and $ W_-(s) $ as functions of $ s $ according
to eq.(\ref{defW}).} 
\label{W}
\end{figure}

Inserting the out of equilibrium distribution eq.(\ref{fuera}) in 
the expression for the DM particle mass eq.(\ref{m})
and using eq.(\ref{disUR}), we obtain the 
generalization of eq.(\ref{mequil}) for the out of LTE case:
\be\label{mfuera}
m = \frac{g_d}{g \; F_0 \; \xi^3} \; \mathrm{eV} \; 
\left\{\begin{array}{l}
         3.593 \; \frac{F_+(\infty)}{F_+(s)}~~~\mathrm{Fermions} \\
         2.695  \; \frac{F_-(\infty)}{F_-(s)}     ~~~\mathrm{Bosons}     
\end{array} \right. \; , 
\ee
where $ s = p_c^0/[\xi \; T_d] $ . Here we used eq.(\ref{disUR}) and 
\be\label{defF}
F_{\pm}(s) \equiv \int_0^s \frac{y^2 \; dy}{e^y \pm 1} \quad , \quad 
F_+(\infty) = \frac32 \; \zeta(3) \quad , \quad F_-(\infty) = 2 \; 
\zeta(3) \quad .
\ee
Inserting the out of equilibrium distribution eq.(\ref{fuera}) into 
eqs.(\ref{mD}) and (\ref{gdD}) for $ m $ and $ g_d $, respectively, 
and using eq.(\ref{disUR}), we obtain the estimates
\bea\label{mgdfeq}
m \sim \left(\frac{Z}{g}\right)^\frac14 \; W_{\pm}(s) \; \mathrm{keV} \;
\left\{\begin{array}{l}
         0.568~~~\mathrm{Fermions} \\
              0.484~~~\mathrm{Bosons}      \end{array} \right. &&
\\ \cr
g_d \sim g^\frac34 \; Z^\frac14 \; \xi^3 \; X_{\pm}(s)
\left\{\begin{array}{l}
         155~~~\mathrm{Fermions} \\
180~~~\mathrm{Bosons}      \end{array} \right. && \label{gdfeq} \; . 
\eea
where
\be\label{defW}
W_{\pm}(s) \equiv \left[\frac{G_{\pm}^3(s) \; F_{\pm}^5(\infty)}{
G_{\pm}^3(\infty) \; F_{\pm}^5(s)}\right]^{\frac18}
\quad , \quad 
X_{\pm}(s) \equiv \left[\frac{G_{\pm}(s) \; 
F_{\pm}(s)}{G_{\pm}(\infty) \; F_{\pm}(\infty)}\right]^{\frac38}
\ee
Here $ F_{\pm}(s) $ is defined by eq.(\ref{defF}) and
\be
G_{\pm}(s) \equiv \int_0^s \frac{y^4 \; dy}{e^y \pm 1}  \; , \;
G_+(\infty) = \frac{45}2 \; \zeta(5) \; , \; 
G_-(\infty) = 24 \; \zeta(5) \; .
\ee
For small arguments $ s $ we have:
$$
W_+(0) = \frac{\sqrt3}{5^\frac34} \; 
\left[\frac{\zeta^5(3)}{\zeta^3(5)}\right]^{\frac18} = 0.5732982\ldots
\; , 
$$
$$
W_-(s) \buildrel{s \to 0}\over= 
\frac{s^\frac14}{2^{\frac58}} \;
\left[\frac{\zeta^5(3)}{27 \; \zeta^3(5)}\right]^{\frac18} =  
0.4753169\ldots \; s^\frac14 \; .
$$
$$
X_+(s) \buildrel{s \to 0}\over= 
\frac{s^3}{\left[2025 \; \zeta(3) \; \zeta(5)\right]^{\frac38}} =
0.0529923\ldots \; s^3 \; , 
$$
$$
X_-(s) \buildrel{s \to 0}\over= 
\frac{s^\frac94}{\left[4320 \; \zeta(3) \; \zeta(5)\right]^{\frac38}} =
0.0398856\ldots \; s^\frac94 \; .
$$
As seen in fig. \ref{W}, 
$$
W_{\pm}(s) \leq 1 \qquad {\rm and} \qquad X_{\pm}(s)\leq 1
\qquad {\rm for} \qquad s \geq 0  \quad  .
$$
We see from eq.(\ref{mgdfeq}) that for relics decoupling out of
LTE, $ m $ is in the {\bf keV range}. 
From eqs.(\ref{mgdfeq})-(\ref{gdfeq}) we see that both 
$ m $ and $ g_d $ for relics decoupling out of LTE
are {\bf smaller} than if they would decouple at LTE.
In addition, since $ X_{\pm}(s) $ vanishes for $ s \to 0 $,
$ g_d $ may be much smaller than for decoupling at LTE.

\medskip

We now generalize eq.(\ref{LTDD}) for the phase-space density 
$ \mathcal{D} $ to the out of LTE case eq.(\ref{LTDD}).
Using eqs.(\ref{Dex}), (\ref{disUR}) and (\ref{fuera}) we have
\be\label{Dfue}
\mathcal{D} = g \; \frac{F_0}{W^4_{\pm}(s)} \; \left\{  \begin{array}{l}
                    1.9625\times 10^{-3}~~\mathrm{Fermions} \\
                          3.6569\times 10^{-3}~~\mathrm{Bosons} 
                                \end{array} \right.
\ee
Inserting now eq.(\ref{Dfue}) into eq.(\ref{meta}) leads to the 
out of LTE generalization of the estimate for the DM particle mass $ m $
eq.(\ref{metae})
\be\label{meta2}
m \sim \left[\frac{\eta \; Z}{g} \; 
\frac{10 \; \frac{\mathrm{km}}{\mathrm{s}}}{\sigma_s}\right]^\frac14 \; 
W_{\pm}(s)
\; \sqrt{\frac{\rm kpc}{L}} \; {\rm keV} \;  \left\{\begin{array}{l}
                    0.405 ~~\mathrm{Fermions} \\
                          0.347 ~~\mathrm{Bosons} 
                                \end{array} \right.
\ee
Taking into account the observed values for $ \sigma_s $ and $ L $ from 
eq.(\ref{gilsL}) and the fact that $ \eta \lesssim 1.6 , \; g \simeq 1-4, 
\; 1 < Z^\frac14 < 10 $, eq.(\ref{meta2}) gives again a mass $ m $ in the
{\bf keV scale} as in eq.(\ref{mgdfeq}). Both equations (\ref{mgdfeq}) and
(\ref{meta2}) yield a mass larger in 17\% for the fermion than for the 
boson.

\subsubsection{An instructive example: Sterile neutrinos decoupling out of LTE.}

We consider in this subsection a sterile neutrino $ \nu $ as DM particle
decoupling out of LTE in a specific model where $ \nu $ is a singlet
Majorana fermion ($ g=2 $) with a Majorana mass $ m_{\nu} $
coupled with a small Yukawa-type coupling 
($ Y \sim 10^{-8} $) to a real scalar field $ \chi $ \citep{sha,gra,jose,misha,otros}. 
$ \chi $ is more strongly coupled to the particles in the Standard Model 
plus to three right-handed neutrinos. As a result, all particles (except 
$ \nu $) remain in LTE well after $ \nu $ decouples from them. 

The distribution function after decoupling of the sterile neutrino $ \nu $
is known for small coupling $ Y $ to be \citep{cinetica},
\be\label{fnu}
F_d^{\nu}(y) = \tau \; \frac{g_{\frac52}(y)}{\sqrt{y}} \qquad {\rm where} 
\qquad g_{\frac52}(y) \equiv \sum_{n=1}^{\infty} 
\frac{e^{-n \, y}}{n^{\frac52}} 
\ee
and the coupling $ \tau $ is in the range $ 0.035 \lesssim \tau \lesssim 
0.35 $ \citep{cinetica}. 

It is interesting to compare the small ($ y \to 0 $) and large momenta 
($ y \to \infty $) behaviour of this out of equilibrium distribution 
$ F_d^{\nu}(y) $ with the Fermi-Dirac equilibrium distribution 
eq.(\ref{disUR}). We find
$$
\frac{F_d^{\nu}(y)}{F_d^{equil}(y)}\buildrel{y \to 0}\over= \frac{2 \; \tau \;
\zeta\left(\frac52\right)}{\sqrt{y}} \to \infty \; , \; 
\zeta\left(\frac52\right)=1.341\ldots   \; ,  
$$
$$
\frac{F_d^{\nu}(y)}{F_d^{equil}(y)}\buildrel{y \to \infty}\over=
\frac{\tau}{\sqrt{y}} \to 0 \; .
$$
Therefore, $ F_d^{\nu}(y) $ exhibits an {\bf enhancement} compared with 
the Fermi-Dirac equilibrium distribution for small ($ y \to 0 $) and a 
{\bf suppression} for large momenta ($ y \to \infty $). Qualitatively, the
out of equilibrium distribution eq.(\ref{fuera}) exhibits the {\bf same 
effect} when compared to the equilibrium distribution $ F_d^{equil} $ as a
 consequence of the incomplete UV cascade.

\medskip

We now evaluate the relevant physical quantities 
inserting $ F_d^{\nu}(y) $ in the relevant equations of sec. \ref{dlte}.
We find for $ m_{\nu} $ from eqs.(\ref{m}) and (\ref{fnu})
\be\label{mnu}
m_{\nu} = 2.34 \; \frac{g_d}{\tau} \; \mathrm{eV} \; ,
\ee
which must be compared with the LTE result for fermions eq.(\ref{mequil}) 
with $ g = 2 $.

The phase-space density $ \mathcal{D} $ from eqs.(\ref{Dex}) and 
(\ref{fnu}) takes the value,
\be\label{Dnu}
\mathcal{D} = \frac{6 \; \tau \; \zeta^{\frac52}(5)}{\left[35 \; \pi 
\; \zeta(7)\right]^{\frac32}} = 5.627 \times 10^{-3} \; \tau 
\quad {\rm where} \quad \zeta(7) = 1.0083493\ldots \; .
\ee
This result is to be compared with the LTE result for fermions 
eq.(\ref{LTDD}) with $ g = 2 $.

Inserting the sterile neutrino distribution function eq.(\ref{fnu}) into
eqs.(\ref{mD}) and (\ref{gdD}), that take into account the decrease of the
phase-space density due to the self-gravity dynamics, we obtain the 
following mass estimates for 
the $ \nu $ DM particles that decoupled out of LTE, 
\be\label{gdnu}
m_{\nu}  \sim \left(\frac{Z}{\tau}\right)^\frac14 \; 0.434 
\; \mathrm{keV} \qquad , \qquad
g_d \sim \tau^\frac34 \; Z^\frac14 \; 185          \; . 
\ee
Again, these formulas must be compared with the LTE result for fermions 
eqs.(\ref{mgdeq}). $ g_d \sim 100 $ corresponds to 
$ T_d \sim 100 $ GeV (see \citet{kt}) which is the expected value for
$ T_d $ in \citet{cinetica}. 

More precisely, for the typical range $  0.035 \lesssim \tau \lesssim 0.35 $,
from eq.(\ref{gdnu}) we find 
$$
0.56 \; \mathrm{keV} \lesssim m_{\nu} \;  Z^{-\frac14} \lesssim 1.0 \; \mathrm{keV}
\qquad ,  \qquad 15 \lesssim g_d  \;  Z^{-\frac14}\lesssim 84 \; ,
$$
while for $ g = 2 $ fermions decoupling in LTE, the mass turns to be smaller:
$ m \;  Z^{-\frac14} = 0.48 $ keV and $ g_d $ larger: $ g_d  \;  Z^{-\frac14} = 184 $ 
[from eq.(\ref{mgdeq})].

A further estimate for $ m_{\nu} $, independent of eq.(\ref{gdnu}), follows by inserting
eq.(\ref{Dnu}) for $ \mathcal{D} $ into eq.(\ref{meta}) valid for a self-gravitating
gas of DM:
\be
m_{\nu} \sim 0.3105 \; \left[\frac{\eta \; Z}{\tau} \; 
\frac{10 \; \mathrm{km}/\mathrm{s}}{\sigma_s}\right]^\frac14 \;
\sqrt{\frac{\rm kpc}{L}} \; {\rm keV} \; ,
\ee
which gives for the typical $\tau$ range,
$$
0.40 \; \mathrm{keV} \lesssim m_{\nu} \;  \left[\eta \; Z \;
\frac{10 \; \mathrm{km}/\mathrm{s}}{\sigma_s}\right]^{-\frac14} \; 
\sqrt{\frac{L}{\rm kpc}} \lesssim 0.72 \; \mathrm{keV}
$$
[Recall that $ 0.1 < Z^{-\frac14} < 1 $.]

\medskip

In summary the results for the sterile neutrino decoupling out of LTE
in the model of \citet{sha,gra,jose,misha,cinetica} are qualitatively
similar to those for fermions decoupling at LTE. 

\section{DM particles decoupling being non-relativistic}\label{secuatro}

Particles decoupling non-relativistic at a temperature $ T_d \ll m $ are 
described by a freezed-out Maxwell-Boltzmann distribution function 
depending on 
$$
\frac{p_c^2}{T_d} = \frac{a^2(t) \; p_{ph}^2(t)}{T_d} = T_d \; y^2 \; .
$$
That is, 
\bea \label{MBf}
&F_d^{equil}(p_c) = \displaystyle
\frac{2^\frac52 \,\pi^\frac72}{45} \; g_d \; 
Y_\infty \; \left(\frac{T_d}{m}\right)^\frac32 \; 
e^{-\frac{p^2_c}{2 \, m \; T_d}} =&\cr \cr
&=\displaystyle\frac{2^\frac52 \,\pi^\frac72}{45} \; g_d \; Y_\infty \;
\left(\frac{T_d}{m}\right)^\frac32 \; e^{- \frac{a^2(t) \; 
p_{ph}^2(t)}{2 \, m \; T_d}}  &\cr \cr
&= \displaystyle\frac{2^\frac52 \,\pi^\frac72}{45} \; 
\frac{g_d \; Y_\infty}{x^\frac32} \;  e^{-\frac{y^2}{2 \; x}} \; ,&
\eea
where $ g_d $ is the effective number of ultrarelativistic degrees of 
freedom at decoupling, $ Y(t)=n(t)/s(t) $,  $ n(t) $ is the number 
of DM particles per unit volume, $ s(t) $ their entropy per unit volume,
$ x \equiv m/T_d $ and $ Y_\infty $ follows from the late time limit of 
the Boltzmann equation \citep{kt,gb}. 

For particles that decoupled NR we obtain inserting eq.(\ref{MBf}) into
the general formula for the DM particle mass eq.(\ref{m}),
\be\label{MNR}
m = \frac{45}{4 \; \pi^2} \; \frac{\Omega_{DM} \;\rho_c}{g \;T_{\gamma}^3
\;  Y_\infty} = \frac{0.748}{g \; Y_\infty} \; {\rm eV} \; . 
\ee
Solving the Boltzmann equation gives for $ Y_\infty $ \citep{kt,gb},
\be\label{yinf}
 Y_\infty = \frac{45}{4 \; \sqrt2 \; \pi^{\frac72}} \; \frac{g}{g_d}
\; x \; e^{-x} \; .
\ee
Notice that $ x \gtrsim 1 $ since the DM particles decoupled NR.
$ Y_\infty $ can also be expressed in terms of $ \sigma_0 $ (the thermally 
averaged total annihilation cross-section times the velocity which appears 
in the Boltzmann equation) as \citep{kt,gb},
\be\label{yinf2}
 Y_\infty = \frac1{\pi} \; \sqrt{\frac{45}8} \; 
\frac{x}{\sqrt{g_d} \; m \; \sigma_0 \; M_{Pl}}
\ee
(We assume for simplicity S-wave annihilation).
It follows from this relation and eq.(\ref{MNR}) that
\be\label{sig0}
\sigma_0 = \frac{0.414 \; 10^{-9}}{\mathrm{GeV}^2} \; 
\frac{g \; x}{\sqrt{g_d}}
\ee
The trascendental equations (\ref{MNR}) and (\ref{yinf}) fix the values 
of $ m $ and $ x $. They can be combined as
\be\label{tras}
 \frac{e^{x}}{x} = 193.5 \; \frac{g^2}{g_d} \; \frac{m}{\rm keV} \; .
\ee
This equation has solutions for $ x > 1 $ provided
$$
\frac{m}{\rm keV} > \frac{e}{193.5} \; \frac{g_d}{g^2} 
= 0.014 \; \frac{g_d}{g^2} \; .
$$
For $ x = m/T_d \gtrsim 1 $ we have the analytic solution of eq.(\ref{tras})
$$
\frac{m}{T_d} = x \simeq \log\left(193.5 \; \frac{g^2}{g_d} \; \frac{m}{\rm keV} \right)
= 5.265 +\log\left(\frac{g^2}{g_d} \; \frac{m}{\rm keV} \right)  \; .
$$
We obtain the mass of the DM particle inserting the non-relativistic
distribution function eq.(\ref{MBf}) into the general formula 
eq.(\ref{mD}) for $ m^4 $ with the result,
\be\label{etam4}
m^\frac52 \; T_d^\frac32 = \frac{45}{2 \; \pi^2} \; 
\frac1{g \; g_d \; Y_\infty} \; Z \; \frac{\rho_s}{\sigma^3_s} \; ,
\ee
Combining eq.(\ref{MNR}) for $ Y_\infty $ with eq.(\ref{etam4}) we obtain for the product 
$ m \; T_d $
\be\label{mTd}
\sqrt{m \; T_d} = 1.47 \; \left(\frac{Z}{g_d}\right)^\frac13 \; 
\mathrm{keV} \;~~~\mathrm{NR~Maxwell-Boltzmann} \; .
\ee
Typical wimps are assumed to have $ m = 100 $ GeV and $ T_d = 5 $ GeV \citep{slac}.
Such value for $ T_d $ implies $ g_d \simeq 80 $ \citep{kt}. Eq.(\ref{mTd}) thus
requires for such heavy wimps $ Z \sim 10^{23} $ well above the upper bounds
derived in sec. \ref{secdos} (see Table I). Therefore, wimps in the 100 GeV scale are
strongly disfavoured. 

\medskip

We find from eqs.(\ref{Dex}) and (\ref{MBf}) the phase-space density
for DM decoupling NR in LTE
\be\label{LTDDNR}
\mathcal{D} = g \; \frac{2 \; \pi^2}{135 \; \sqrt3} \; g_d \; Y_\infty 
\; \left(\frac{T_d}{m}\right)^\frac32 =  8.4418 \times 10^{-2} \; g \; 
g_d \; Y_\infty \; x^{-\frac32} \; .
\ee
We obtain using here the value for $ Y_\infty $ from eq.(\ref{MNR}) 
\be\label{DNR}
\mathcal{D} = 0.6315 \; 10^{-4} \; g_d \; \frac{\mathrm{keV}}{m^\frac52}
\; T_d^\frac32 \; .
\ee
Inserting this expression for $ \mathcal{D} $ into the general estimate
for the DM mass eq.(\ref{meta}) yields,
\be
\sqrt{m \; T_d} \sim 0.942 \; \left(\frac{\eta \; Z}{g_d}\right)^\frac13
\; \left(\frac{\rm kpc}{L}\right)^\frac23 \; 
\left(\frac{10\;\mathrm{km}/\mathrm{s}}{\sigma_s}\right)^\frac13
\; {\rm keV} \; .
\ee
As in eq.(\ref{mTd}) but independently from it,
we reach a result for $ \sqrt{m \; T_d} $ in the keV scale assuming the DM
is a self-gravitating gas in thermal equilibrium.

\subsection{Allowed ranges for $ m, \;  T_d $ and the annihilation cross section $ \sigma_0 $.}

We derive here individual bounds on $ m, \;  T_d $ and $ \sigma_0 $ for DM
particles decoupling NR.

\noindent
Using that $ T_d < m $ for DM particles that decoupled NR we obtain
from eq.(\ref{mTd}) a lower bound for $ m $ and an upper bound on 
$ T_d $. Furthermore, taking into account that
$ T_d > b $ eV where $ b > 1 $ or $ b \gg 1 $ for DM particles that decoupled
in the RD era, we obtain an upper bound for $ m $. In summary,
\bea\label{cotamNR}
&&\left(\frac{Z}{g_d}\right)^\frac13 \; 1.47 \; 
\mathrm{keV} < m < \frac{2.16}{b} \; \mathrm{MeV} \; 
\left(\frac{Z}{g_d}\right)^\frac23  \; , \cr\cr
&& b \; \mathrm{eV} < T_d < \left(\frac{Z}{g_d}\right)^\frac13 \; 
1.47 \; \mathrm{keV} \; . 
\eea
Recalling that \citep{kt}
\be\label{gd3}
g_d \simeq 3 \quad {\rm for} \quad 1 ~ \mathrm{eV} < T_d < 100 ~ 
\mathrm{keV}  \; ,
\ee
and that  $ 1 < Z < 10^4 $, we see from eqs.(\ref{cotamNR}) that 
$$
1.02 \; \mathrm{keV} < m < \frac{482}{b} \; \mathrm{MeV} 
\quad , \quad T_d < 10.2 \; \mathrm{keV} \; .
$$
Notice that $ b $ may be {\bf much larger} than one 
with $ b < 1470 \; \left(Z/g_d\right)^\frac13  < 21960 $ to ensure
$ T_d < m $ for consistency in eqs.(\ref{cotamNR}).

\medskip

In addition, lower and upper bounds for the cross-section $ \sigma_0 $ 
can be derived. From eqs.(\ref{sig0}), (\ref{gd3}) and $ x > 1 $ a lower
bound follows,
\be\label{sigco}
\sigma_0 > 0.239 \; 10^{-9} \; \mathrm{GeV}^{-2} \; g \; .
\ee
On the other hand, upper bounds for the total DM self-interaction cross sections 
$ \sigma_T $ have been given by comparing X-ray, optical and lensing observations 
of the merging of galaxy clusters with $N$-body simulations in \citet{clowe,randall,brada} 
(see also \citet{ante,ostri,arab}):
$$
\frac{\sigma_T}{m} < 0.7 \; \frac{{\rm cm}^2}{\rm gr} = 3200 \; \mathrm{GeV}^{-3} \; .
$$
Since the annihilation cross-section must
be smaller than the total cross-section we can write the bound
\be
\sigma_0 < 3200 \; m \; \mathrm{GeV}^{-3} \; .
\ee
Using the upper bound eq.(\ref{cotamNR}) for $ m $ yields
the upper bound for $ \sigma_0 $
\be
\sigma_0 < \frac{3.32 \; Z^\frac23}{b} \; \mathrm{GeV}^{-2} \; .
\ee
This result leaves at least five orders of magnitude between the lower 
bound [eq.(\ref{sigco})] and the upper bound for $ \sigma_0 $. The DM 
non-gravitational self-interaction is therefore negligible in this 
context.

\medskip

Exotic models where very heavy ($ \sim 10 $ TeV) DM particles are produced
very late and decouple non-relativistically were proposed 
introducing two new fine tuned parameters: (a) the lifetime of unstable 
particles (sneutrinos) that decay into DM (gravitinos) 
(b) the mass difference between the two particles which must be small 
enough to led to non-relativistic DM \citep{cfrt,skb}.
It is stated in \cite{cfrt} and \cite{skb} that such models may describe 
the observed phase-space density. It is stated in \cite{bbu} that
it is inherently difficult to fulfil all observational constraints
in such models.

\section{Conclusions}

Our results are {\bf independent} of the particle model that will 
describe the dark matter. We consider both DM particles that decouple 
being NR and UR and both decoupling at LTE and out of LTE.
Our analysis and results refer to the mass of the dark matter particle 
and the number of ultrarelativistic effective degrees of freedom when the
DM particles decoupled. We do not make assumptions about the nature of 
the DM particle and we only assume that its non-gravitational 
interactions can be neglected in the present context (which is
consistent with structure formation and observations).

In case DM particles explain the formation of galactic
center black holes, DM particles must be fermions with keV-scale mass
\citep{mubi}.

\noindent

The mass for the DM particle in the keV range is much larger
than the temperature during the MD era, hence dark matter is {\bf cold} 
(CDM).

A possible CDM candidate in the keV scale is a sterile neutrino 
\citep{esteril,shifu,afp,aba,mubi,kuse} produced via their mixing and oscillation 
with an active 
neutrino species. Other putative CDM candidates in the keV scale are the 
gravitino \citep{gravitino,Steffen}, the light neutralino  \citep{neutr} 
and the majoron \citep{valle}. 

Actually, many more extensions of the Standard Model
of Particle Physics can be envisaged to
include a DM particle with mass in the keV scale and
weakly enough coupled to the Standard Model particles.

Lyman-$\alpha$ forest observations provide indirect lower bounds on the 
masses of sterile neutrinos \citep{lyalf,lyalf2} while constraints from the 
diffuse X-ray background yield upper bounds on the mass of a putative 
sterile neutrino DM particle \citep{rX,watson,abo,rs,rs2,lkb}. All these recent constraints are 
consistent with DM particle masses at the keV scale.

The DAMA/LIBRA collaboration has confirmed the presence of a signal in the
keV range \citep{dama}. Whether this signal is due to DM particles
in the keV mass scale is still unclear \citep{contra,contra2,prv}. On the other hand,
the DAMA/LIBRA signals cannot be explained
by a hypothetical WIMP particle with mass $ \gtrsim O(1) $ GeV since
this would be in conflict with previous WIMPS direct detection experiments 
\citep{wimp,fasch,hoop,sava,sadou}.

\medskip

As discussed in sec. \ref{secuatro}, typical wimps with $ m = 100 $ GeV 
and $ T_d = 5 $ GeV  \citep{slac} would require a huge $ Z \sim 10^{23} $, well above
the upper bounds displayed in Table I. Hence, wimps cannot reproduce
the observed galaxy properties. In addition, recall that $ Z \sim 10^{23} $ produces
from eq.(\ref{fs2}) an extremely short $ \lambda_{fs} $ today 
$$
\lambda_{fs}(0) \sim 3.51 \; 10^{-4} \; {\rm pc} = 72.4  \; {\rm AU} \; .
$$

If the flyby anomaly would be explained by DM, a keV scale DM mass is
preferred \citep{adler}.

Further evidence for the DM particle mass in the keV scale follows by contrasting
the observed value of the constant surface density of galaxies 
to the theoretical calculation from the linearized Boltzmann-Vlasov equation \citep{dsg}.
Independent further evidence for the DM particle mass in the keV scale
is given by \citet{tikho}. [See also \citet{gilmore2}].

\medskip

In summary, our analysis shows that DM particles decoupling UR in LTE have
a mass $ m $ in the keV scale with $ g_d \gtrsim 150 $  as shown in 
sec. \ref{dlte}. That is, decoupling happens at least at the 100 GeV 
scale. The values of $ m $ and $ g_d $ may be smaller for DM decoupling UR
out of LTE than for decoupling UR in LTE (see sec. \ref{dflte}). For DM 
particles decoupling NR in LTE ($ T_d < m $) we find that 
$ \sqrt{m \; T_d} $ is in the keV range. This is consistent with the DM 
particle mass in the keV range. 

Notice that the present uncertainity by one order of magnitude
of the observed values of the phase-space density $ \rho_s/\sigma^3_s $ only 
affects the DM particle mass through a power $ 1/4 $ of this uncertainity according to 
eqs. (\ref{mD})-(\ref{mDn}). Namely, by a factor $ 10^{\frac14} \simeq 1.8 $.

We find that the free streaming wavelength (Jeans' length) is 
{\bf independent} of the nature of the DM particle except for the $ Z $ 
factor characterizing the decrease of the phase-space density through  
self-gravity [sec. \ref{jeans}]. The values found for the Jeans' length 
and the Jeans' mass for $ m $ in the keV scale 
are consistent with the observed small structure and 
with the masses of the galaxies, respectively.

\section*{acknowledgments}
We thank C. Alard, D. Boyanovsky, C. Frenk, G. Gilmore, B. Sadoulet, P. Salucci 
for fruitful discussions.

\label{lastpage}

\begin{thebibliography}{99}
\bibitem[\protect\citeauthoryear{Aalseth et al.}{2008}]{wimp} 
C. E. Aalseth et al. Phys. Rev. Lett. 101, 251301 (2008).

\bibitem[\protect\citeauthoryear{Abazajian, Fuller \& Patel}{2001}]{afp}
K. Abazajian, G. M. Fuller, M. Patel, Phys. Rev. \textbf{D64}, 023501 (2001).

\bibitem[\protect\citeauthoryear{Abazajian}{2006}]{aba}
K. Abazajian, Phys. Rev. \textbf{D73}, 063506 (2006).

\bibitem[\protect\citeauthoryear{Adler}{2009}]{adler} 
S. L. Adler, Phys. Rev. D79, 023505 (2009).

\bibitem[\protect\citeauthoryear{Ahmed et al.}{2009}]{sadou}
Z. Ahmed et al., arXiv:0907.1438.

\bibitem[\protect\citeauthoryear{Arabadjis et al.}{2002}]{arab}
J. S. Arabadjis et al. Ap. J. 572, 66 (2002).

\bibitem[\protect\citeauthoryear{Bernabei et al.}{2006}]{contra}
R. Bernabei et al. Int. J. Mod. Phys. A21, 1445 (2006).

\bibitem[\protect\citeauthoryear{Bernabei et al.}{2008a}]{dama} 
R. Bernabei et al. (DAMA/LIBRA coll.) Eur. Phys. J. C56:333 (2008a). 

\bibitem[\protect\citeauthoryear{Bernabei et al.}{2008b}]{contra2} 
R. Bernabei et al. Mod. Phys. Lett. A23:2125 (2008b).

\bibitem[\protect\citeauthoryear{Bertschinger}{1985}]{EB}
E. Bertschinger, ApJS, 58, 1 and 39 (1985).

\bibitem[\protect\citeauthoryear{Binney \& Tremaine}{1987}]{bt} 
J. Binney, S. Tremaine, \emph{Galactic Dynamics},
Princeton University Press, 1987.

\bibitem[\protect\citeauthoryear{Bond, Szalay \& Turner}{1982}]{bst}
J R Bond, A S Szalay, M S Turner, Phys. Rev. Lett. \textbf{48}, 1636 (1982).

\bibitem[\protect\citeauthoryear{Bond \& Szalay}{1983}]{bs} 
J R Bond, A S Szalay, Astrophys. J. {\bf 274}, 443 (1983).

\bibitem[\protect\citeauthoryear{B\"orner}{2003}]{gb} 
G. B\"orner, \emph{The Early Universe}, Springer, 2003.

\bibitem[\protect\citeauthoryear{Borzumati et al.}{2008}]{bbu}
F. Borzumati et al., Phys. Rev. D77, 063514 (2008).

\bibitem[\protect\citeauthoryear{Boyanovsky, Destri \& de Vega}{2004}]{DdV}  
D. Boyanovsky, C. Destri, H. J. de Vega, Phys. Rev. {\bf D 69}, 045003 (2004).

\bibitem[\protect\citeauthoryear{Boyanovsky, de Vega \& Sanchez}{2008a}]{nos1} 
D. Boyanovsky, H. J. de Vega, N. Sanchez,
Phys. Rev. {\bf D 77}, 043518 (2008a), arXiv:0710.5180.

\bibitem[\protect\citeauthoryear{Boyanovsky, de Vega \& Sanchez}{2008b}]{egil} 
D. Boyanovsky, H. J. de Vega, N. Sanchez, Phys. Rev. {\bf D 78}, 063546 (2008b).

\bibitem[\protect\citeauthoryear{Boyanovsky}{2008}]{cinetica} 
D. Boyanovsky, Phys. Rev. D78:103505, (2008).

\bibitem[\protect\citeauthoryear{Boyarsky et al.}{2007}]{abo}
A. Boyarsky et al. Astron. Astrophys. 471:51 (2007).

\bibitem[\protect\citeauthoryear{Brada}{2008}]{brada}
M. Brada$\check{\rm c}$ et al. Ap. J. 687, 959 (2008).

\bibitem[\protect\citeauthoryear{Cembranos et al.}{2005}]{cfrt}
J Cembranos  et al. Phys. Rev. Lett. \textbf{95}, 181301 (2005).

\bibitem[\protect\citeauthoryear{Chikashige et al.}{1981}]{sha} 
Y. Chikashige, R. N. Mohapatra, R. D. Peccei, Phys. Lett. 98B, 
265 (1981).

\bibitem[\protect\citeauthoryear{Cowsick \& McClelland}{1972}]{cow} 
R. Cowsick and J. McClelland, Phys. Rev. Lett. \textbf{29}, 669 (1972).

\bibitem[\protect\citeauthoryear{Destri \& de Vega}{2006}]{DdV2}  
C. Destri, H. J. de Vega, Phys. Rev. {\bf D 73}, 025014 (2006).

\bibitem[\protect\citeauthoryear{Destri \& de Vega}{2007}]{gas2} 
C. Destri, H. J. de Vega, Nucl. Phys. {\bf B 763}, 309 (2007).

\bibitem[\protect\citeauthoryear{de Vega \& S\'anchez}{2002}]{gas} 
H. J. de Vega, N. S\'anchez, Nucl. Phys. {\bf B 625}, 409 and 460 (2002).

\bibitem[\protect\citeauthoryear{de Vega \& S\'anchez}{2009}]{dsg}
H. J. de Vega, N. S\'anchez, arXiv:0907.0006.

\bibitem[\protect\citeauthoryear{Dodelson}{2003}]{dod}
Dodelson S, \textit{Modern Cosmology},  Academic Press, 2003.

\bibitem[\protect\citeauthoryear{Dodelson \& Widrow}{1994}]{esteril}  
S. Dodelson, L. M. Widrow, Phys. Rev. Lett. \textbf{72}, 17 (1994).

\bibitem[\protect\citeauthoryear{Dolgov \& Hansen}{2002}]{rX} 
A. D. Dolgov and S.  H. Hansen, Astropart. Phys. \textbf{16}, 339 (2002).

\bibitem[\protect\citeauthoryear{Fairbairn \& Schwetz}{2009}]{fasch}
M. Fairbairn, T. Schwetz, JCAP 0901:037, (2009).

\bibitem[\protect\citeauthoryear{Fillmore \& Goldrich}{1984}]{FG}
J. A. Fillmore, P. Goldrich, ApJ, 281, 1 and 9 (1984).

\bibitem[\protect\citeauthoryear{Gao \& Theuns}{2007}]{gao} 
L. Gao, T. Theuns, Science 317:1527 (2007). 

\bibitem[\protect\citeauthoryear{Gelmini \& Roncadelli}{1981}]{gra}
G. B. Gelmini, M. Roncadelli, Phys. Lett. 99B, 411 (1981).

\bibitem[\protect\citeauthoryear{Gilbert}{1968}]{gilbert} 
I. H. Gilbert, Astrophys. J. \textbf{144}, 233 (1966); \emph{ibid}, \textbf{152}, 1043 (1968).

\bibitem[\protect\citeauthoryear{Gilmore et. al.}{2007}]{gilmore2} 
G. Gilmore \emph{et. al.} Astrophys. J, 663, 948 (2007).

\bibitem[\protect\citeauthoryear{Gorbunov et al.}{2008}]{gravitino} 
D. Gorbunov, A. Khmelnitsky, V. Rubakov, JHEP 0812:055 (2008).

\bibitem[\protect\citeauthoryear{Hennawi \& Ostriker}{2002}]{ostri}
J. F. Hennawi, J. P. Ostriker,  Ap. J. 572, 41 (2002).

\bibitem[\protect\citeauthoryear{Hoffman et al.}{2007}]{numQ2} 
Y. Hoffman, E. Romano-Diaz, I. Shlosman, C. Heller,
Astrophys. J. {\bf 671}, 1108 (2007). 

\bibitem[\protect\citeauthoryear{Hogan \& Dalcanton}{2000}]{hogan} 
C. J. Hogan, J. J. Dalcanton, Phys. Rev. \textbf{D62}, 063511 (2000), 
J. J. Dalcanton, C. J. Hogan, Astrophys. J.\textbf{561}, 35 (2001).

\bibitem[\protect\citeauthoryear{Hooper et al.}{2009}]{hoop}
D. Hooper et al. Phys. Rev. D79:015010 (2009).

\bibitem[\protect\citeauthoryear{Kolb \& Turner}{1990}]{kt} 
E. W. Kolb, M. S. Turner, \emph{The Early Universe}, Addison-Wesley (1990).

\bibitem[\protect\citeauthoryear{Komatsu et al.}{2009}]{WMAP5} 
E. Komatsu et al. (WMAP collaboration), Astrophys. J. Suppl. 180:330 (2009).

\bibitem[\protect\citeauthoryear{Kusenko}{2007}]{kuse}
A. Kusenko, Int. J. Mod. Phys. D16:2325, (2007).

\bibitem[\protect\citeauthoryear{Lapi \& Cavaliere}{2009}]{numQ3} 
A. Lapi, A. Cavaliere, Astrophys. J. {\bf 692},  1, 174 (2009).

\bibitem[\protect\citeauthoryear{Lattanzi \& Valle}{2007}]{valle} 
M. Lattanzi, J.W.F. Valle, Phys. Rev. Lett. 99:121301 (2007). 

\bibitem[\protect\citeauthoryear{Lee \& Weinberg}{1977}]{LW} 
B W Lee, S. Weinberg, Phys. Rev. Lett. \textbf{39}, 165 (1977).

\bibitem[\protect\citeauthoryear{Loewenstein et al.}{2009}]{lkb}
M. Loewenstein, A. Kusenko, P. L. Biermann, Astrophys. J. 700:426-435 (2009).

\bibitem[\protect\citeauthoryear{Lynden-Bell}{1967}]{theo} 
D. Lynden-Bell, Mon. Not. Roy. Astron. Soc. \textbf{136}, 101 (1967).

\bibitem[\protect\citeauthoryear{Madsen}{1990}]{madsen} 
J. Madsen, Phys. Rev. Lett. \textbf{64}, 2744 (1990).

\bibitem[\protect\citeauthoryear{Madsen}{2001}]{madsen2} 
J. Madsen, Phys. Rev. \textbf{D64}, 027301 (2001).

\bibitem[\protect\citeauthoryear{Markevitch et al.}{2004}]{clowe} 
M. Markevitch et al. Ap. J. 606, 819 (2004). 

\bibitem[\protect\citeauthoryear{McDonald \& Sahu}{2009}]{otros} 
J. McDonald, N. Sahu, 	Phys. Rev. D 79, 103523 (2009).

\bibitem[\protect\citeauthoryear{Miralda-Escud\'e}{2002}]{ante} 
J. Miralda-Escud\'e, Ap. J. 564, 60 (2002).

\bibitem[\protect\citeauthoryear{Munyaneza \& Biermann}{2006}]{mubi}
F. Munyaneza, P. L. Biermann, Astron. and Astrophys., 458, L9 (2006).

\bibitem[\protect\citeauthoryear{Oort}{1940}]{oo}
J. H. Oort, ApJ, 91, 273 (1940).
See S. van den Bergh, astro-ph/0005314 for a history of the research 
on dark matter.

\bibitem[\protect\citeauthoryear{Padmanabhan}{1999}]{TP}
T. Padmanabhan, astro-ph/9911374, Lectures at the IPM School,
Kluwer, Dordrecht, 2000.

\bibitem[\protect\citeauthoryear{Pagels \& Primack}{1982}]{bstpp} 
H. Pagels, J R Primack, Phys. Rev. Lett. \textbf{48}, 223 (1982). 

\bibitem[\protect\citeauthoryear{Particle Data Group}{2009}]{slac}
Particle Data Group, http://pdg.lbl.gov

\bibitem[\protect\citeauthoryear{Peebles}{1993}]{PJEP}
P. J. E. Peebles, Principles of Physical Cosmology, Princeton Univ. Press,
Princeton NJ, 1993.

\bibitem[\protect\citeauthoryear{Peirani et al.}{2006}]{numQ1} 
S. Peirani \emph{et. al.}, Mon. Not. R. Astron. Soc.
\textbf{367}, 1011 (2006).

\bibitem[\protect\citeauthoryear{Pospelov et al.}{2008}]{prv}
M. Pospelov, A. Ritz, M. B. Voloshin, Phys. Rev. D78:115012 (2008).

\bibitem[\protect\citeauthoryear{Profumo}{2008}]{neutr} 
S. Profumo, Phys. Rev. D78:023507 (2008) and references therein.

\bibitem[\protect\citeauthoryear{Randall et al.}{2008}]{randall}
S. W. Randall  et al. Ap. J. 679, 1173 (2008).

\bibitem[\protect\citeauthoryear{Riemer-Sorensen et al.}{2006}]{rs}
S.~Riemer-Sorensen et al. Astrophys.\ J.\  {\bf 644}, L33 (2006).

\bibitem[\protect\citeauthoryear{Riemer-Sorensen et al.}{2007}]{rs2}
S.~Riemer-Sorensen et al. Phys. Rev. D76:043524, (2007).

\bibitem[\protect\citeauthoryear{Romano-Diaz et al.}{2006}]{numQ4} 
E. Romano-Diaz \emph{et.al.}, Astrophys. J. \textbf{637}, L93 (2006). 

\bibitem[\protect\citeauthoryear{Romano-Diaz et al.}{2007}]{numQ5} 
E. Romano-Diaz \emph{et.al.}, Astrophys. J. \textbf{657}, 56 (2007).
 
\bibitem[\protect\citeauthoryear{Sato \& Kobayashi}{1977}]{SK}
K. Sato, H. Kobayashi, Prog. Theor. Phys. \textbf{58}, 1775 (1977). 

\bibitem[\protect\citeauthoryear{Savage et al.}{2008}]{sava}
C. Savage et al. arXiv:0808.3607.

\bibitem[\protect\citeauthoryear{Schechter \& Valle}{1982}]{jose}
J. Schechter, J.W.F. Valle, Phys. Rev. D25, 774 (1982).

\bibitem[\protect\citeauthoryear{Shaposhnikov \& Tkachev}{2006}]{misha}
M. Shaposhnikov, I. Tkachev, Phys. Lett. B639, 414 (2006).

\bibitem[\protect\citeauthoryear{Shi \& Fuller}{1999}]{shifu}
X. Shi, G. M. Fuller, Phys. Rev. Lett. \textbf{82}, 2832 (1999).

\bibitem[\protect\citeauthoryear{Steffen}{2009}]{Steffen}
F. D. Steffen, Eur. Phys. J. C59, 557 (2009) and references therein.

\bibitem[\protect\citeauthoryear{Strigari et al.}{2007}]{skb}
L. E. Strigari  et al. Phys. Rev. D75, 061303 (2007).

\bibitem[\protect\citeauthoryear{Tikhonov et al.}{2009}]{tikho}
A. V. Tikhonov et al. arXiv:0904.0175, to be published in MNRAS.

\bibitem[\protect\citeauthoryear{Tremaine et al.}{1986}]{theo2} 
S. Tremaine, M. Henon, D. Lynden-Bell, Mon. Not. Roy. Astron. Soc.
\textbf{219}, 285 (1986).

\bibitem[\protect\citeauthoryear{Vass et al.}{2009}]{numQ6}
I. M. Vass et al.,  Mon. Not. R. Astron. Soc. {\bf 395}, 1225 (2009).

\bibitem[\protect\citeauthoryear{Viel et al.}{2005}]{lyalf} 
M. Viel \emph{et.al.} Phys. Rev. \textbf{D71}, 063534 (2005).

\bibitem[\protect\citeauthoryear{Viel et al.}{2007}]{lyalf2} 
M. Viel \emph{et.al.} Phys. Rev. Lett. 100:041304 (2007).

\bibitem[\protect\citeauthoryear{Vysotsky, Dolgov \& Zeldovich}{1977}]{VDZ}
M. I. Vysotsky, A. D. Dolgov, Ya. B. Zeldovich, JETP Lett.\textbf{ 26}, 188 (1977).

\bibitem[\protect\citeauthoryear{Watson et al.}{2006}]{watson}
C. R. Watson et al., Phys. Rev. \textbf{D74}, 033009 (2006). 

\bibitem[\protect\citeauthoryear{Wyse \& Gilmore}{2007}]{gilmore} 
R. F. G. Wyse and  G. Gilmore, arXiv:0708.1492; 

\bibitem[\protect\citeauthoryear{Yao}{2006}]{pdg} W.-M. Yao et al., 
Journal of Physics G 33, 1 (2006).

\bibitem[\protect\citeauthoryear{Zwicky}{1933}]{zw} 
F. Zwicky, Helv. Phys. Acta, 6, 124 (1933).

\end{thebibliography}
\end{document}